**Polytropic Behavior in Corotating Interaction Regions: Evidence of Alfvénic Heating**

M. A. Dayeh[1,2], M. J. Starkey[1], G. Livadiotis[3], S. Hart[1], A. A. Shmies[2], R. C. Allen[1], R. Bučík[1], and H. Elliott[1]

[1] *Southwest Research Institute, San Antonio, TX 78238 (maldayeh@swri.org)*

[2] *University of Texas at San Antonio, San Antonio, TX 78249*

[3] *Princeton University, Princeton, NJ, 08544*

**Abstract.** Corotating Interaction Regions (CIRs) are recurring structures in the solar wind, characterized by interactions between fast and slow solar wind streams that compress and heat plasma. This study investigates the polytropic behavior of distinct regions in and around CIRs: uncompressed slow solar wind, compressed slow solar wind, compressed fast solar wind, and uncompressed fast solar wind. Using *Wind* spacecraft data and an established methodology for calculating the polytropic index ($\gamma$), we analyze 117 CIR events. Results indicate varying $\gamma$ values across regions, with heating observed in compressed regions driven by Alfvén wave dissipation originating from fast streams. In the uncompressed fast solar wind, $\gamma$ exceeds adiabatic values the most and correlates well with strong Alfvénic wave activity.

# 1. Introduction

Corotating Interaction Regions (CIRs) are large-scale structures in the solar wind where fast streams interact with slower streams emitted earlier, producing compression and heating of the plasma. These regions are characterized by distinct boundaries, known as stream interfaces, which are identified by a clear decrease in plasma density, an increase in temperature, and a small jump in the bulk flow speed (Newbury et al. 1997, Tu et al. 1990, Gosling and Pizzo 1999, Borovsky et al., 2006). CIRs have been observed as close as 0.17 au (Allen et al., 2021) out to Jupiter's orbit and beyond (e.g., Van Hollebeke et al., 1978), evolving as they propagate outward (e.g., Gazis 2000). At much greater distances, near the heliospheric boundary, multiple CIRs and other solar transients (shocks, ejecta) merge to form Merged Interaction Regions (MIR), which play a key role in shaping the large-scale structure of the outer heliosphere. By comparing Figures 2 and 3 in Elliott et al. (2019) it is apparent that 10-15 solar wind structures at 1 au merge to produce one larger lower amplitude structure beyond 20 au. CIRs play a significant role in space weather in the near-Earth environment. They can drive strong enhancements in solar wind dynamic pressure, sometimes even driving shocks near 1 au, that can notably compress the dayside magnetopause and facilitate gradual energy coupling to the magnetosphere, resulting in prolonged geomagnetic activity (e.g., Chen et al. 2012; Allen et al., 2023). This sustained interaction can lead to considerable energy transfer into the Earth's magnetosphere, impacting numerous space weather phenomena including auroral oval dynamics and ring current growth and evolution (e.g., Perreault and Akasofu 1978; Allen et al., 2023), ionospheric/thermospheric temperature, density, collision frequencies, upwelling (Allen et al., 2023), and increasing risks of spacecraft charging (e.g., Miteva et al. 2023). Kp can reach as high as 6 and 7 in CIRs owing to the high speeds and densities in the CIRs compressions (Elliott et al., 2013). Turbulence is a key feature of CIRs as it is embedded in both the slow and fast interacting solar wind streams. Slow wind typically shows well-developed turbulence with a near-Kolmogorov spectrum, strong density fluctuations, and low Alfvénicity (Klein 1993, Bruno 1997). In contrast, fast wind is characterized by high Alfvénicity, outward-propagating Alfvén waves, and a flatter energy spectrum (Bruno 1997, Horbury and Schmidt 1999, D'Amicis et al. 2022). Developed CIRs exhibit complex turbulent signatures, with evidence of plasma mixing and possible instabilities at stream interfaces (Gulamali and Cargill, 2001). A superposed-epoch analysis of CIRs showed smooth variations in turbulence characteristics across its structure, including fluctuation amplitudes, Alfvénicity, and spectral slopes, with no clear evidence of turbulence driving by shear at the stream interface (Borovsky and Denton, 2010).

Understanding the thermodynamic behavior within CIRs is essential for explaining the energy transfer processes governing their evolution. A key parameter is the polytropic index (γ), which defines the relationship between plasma moments, that is, temperature *T*, density *n*, and thermal pressure *P*, during compression or expansion (e.g., Totten et al., 1995; Nicolaou et al., 2014; Livadiotis and Desai, 2016). In a simplified form, the polytropic relationship can be expressed mathematically as:

$$P \cdot n^{-\gamma} = T \cdot n^{1-\gamma} = const., \qquad (1)$$

where the constant labels different stationary states and involves another thermodynamic parameter called polytropic pressure Π (Livadiotis 2016, Livadiotis et al. 2022); the values of Π together with the values of any parameter of the triplet (*n*, *T*, *P*) form a grid that describes the thermodynamic phase-space equivalently to the grid of any two original thermodynamic parameters (typically, *P* and *n*).

The polytropic expression for the logarithms of the thermal variables in equation (1) become linear and allows a straightforward estimation of *γ* by simply examining datasets of log(*T*) vs. log(*n*) and performing a linear regression. This is statistically meaningful because space plasma moments have values described by log-normal distributions in general (thus their logarithms are normally distributed). The γ parameter informs how plasma behaves through alternative thermodynamic pathways: γ=∞ for isochoric (constant volume), γ=5/3 for adiabatic (no heat transfer), γ=1 for isothermal (constant temperature), γ=0 for isobaric (constant pressure), and negative γ indicates explosive behavior, (for a review, see Livadiotis and McComas 2012). Importance of the γ parameter in space plasma studies extends beyond CIRs, as it provides a versatile diagnostic of plasma thermodynamics in diverse solar wind environments, including solar wind (e.g., Nicolaou et al. 2014; Livadiotis & Desai 2016; Livadiotis 2018; Livadiotis & Nicolaou 2021; Livadiotis & McComas 2023a,b; Katsavrias et al. 2024), coronal mass ejections (e.g., Dayeh et al. 2022), and interactions of solar disturbances (e.g., Ghag et al., 2025).

Localized variations in the polytropic index enable studying the thermodynamic behavior of interacting plasma structures (Totten et al., 1995). For instance, Newbury et al. (1997) explored polytropic behavior in the vicinity of stream interfaces, showing how solar corona inhomogeneities influence large-scale applications of the polytropic law. Livadiotis (2018) studied long-term trends in the solar wind's γ, particularly its variations and independence from flow speed. The authors demonstrated that γ remains quasi-constant near Earth's orbit (1 au), irrespective of solar wind speed. The largest fluctuations were observed in the fast solar wind, while the γ 's stability analysis indicated consistent thermodynamic states over time. Katsavrias et al. (2024) demonstrated that coherent periodic plasma density structures exhibit sub-adiabatic γ values and higher entropy compared to the overall Wind dataset, while incoherent structures align with dataset-wide γ values. Finally, Dayeh et al. (2022) analyzed polytropic trends in 336 interplanetary coronal mass ejections (ICMEs), revealing that sheath regions exhibit the lowest γ and highest turbulence levels, whereas ejecta regions are sub-adiabatic but less turbulent, surrounded by adiabatic solar wind.

In this study, we investigate the polytropic behavior in four distinct regions in and around CIRs, namely: unperturbed slow solar wind, compressed slow solar wind, compressed fast solar wind, and uncompressed fast solar wind. Our findings reveal evidence of heating in compressed solar wind regions, likely driven by Alfvén wave dissipation originating from fast solar wind streams.

## 2. Data

Events analyzed in this study are provided by the Broiles et al. (2012) catalogue of CIRs observed by the Wind spacecraft from January, 1995 through December, 2008. The catalogue comprises 153 events which were identified by an automated algorithm using various CIR-associated plasma and magnetic field parameters (e.g., proton speed, proton density, magnetic field strength, etc.). For each event in the catalogue, the following boundaries were also identified: stream interface (SI), forward and reverse boundaries (FB and RB, respectively) of the CIR, and the end of the high-speed stream (HSS). We use these identified boundaries to define four distinct regions of the CIR:

1) Uncompressed slow solar wind (SSW) spanning 1 day before the FB;
2) Compressed SSW from the FB to the SI;
3) Compressed fast solar wind (FSW) from the SI to the RB;
4) Uncompressed FSW from the RB to HSS.

We use high cadence (96 seconds) solar wind plasma moments and magnetic field measurements from the Solar Wind Experiment (SWE; Ogilvie et al. 1995) and the Magnetic Field Investigation (Lepping et al. 1995) onboard the Wind spacecraft, respectively. All data is publicly available on the coordinated Data Analysis Web (CDAWeb). The solar wind proton moments are derived using the Levenberg-Marquardt non-linear least-squares fitting method on the bi-Maxwellian velocity distribution functions observed by the SWE Faraday cup (see Kasper 2002 for details).

Two 'product types' of temperature and density datasets are publicly available; one is derived from statistical moments and the other from bi-Maxwellian fitting. The first type is more accurate because it does not include the systematic errors that originate from using Maxwellian instead of kappa distributions. On the other hand, the second type is less erroneous, because the performed fitting smooths noisy effects. As Nicolaou & Livadiotis (2016) had shown, for the polytropic index investigations, it is advised to use the bi-Maxwellian fitting type, because the systematic error is less effective than the noise of the moment type of datasets.

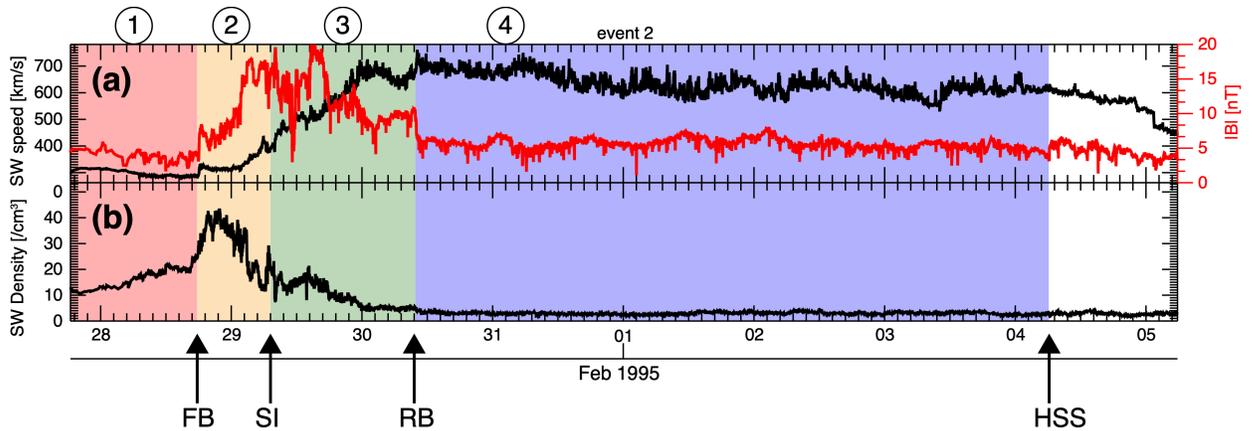

*Figure 1: (a) Solar wind speed (black) and magnetic field strength (red) corresponding to Event 2 (1995-01-27) from the catalogue used (Broiles et al. 2012). (b) Solar wind proton density. The four boundaries are indicated at the bottom of the plot and the corresponding regions of the CIR are indicated by color.*

Figure 1 shows an overview of one selected event, with the four distinct CIR regions indicated by color and numerically labelled on top (1 through 4). Panel (a) shows the solar wind speed (black) and magnetic field strength (red), and panel (b) shows the proton density. Selection criteria of these regions are detailed in Broiles et al. (2012). It is worth noting that different criteria yields slightly different boundaries for some events when the change is not sharp. From the 153 events, we select a subset of 148 events where each region spans at least 20 minutes ensuring sufficient statistics for the polytropic index calculations.

## 3. Methodology

**Calculation of the polytropic index**

We aim at determining the polytropic state in each of the subregions defined in Figure 1 for all selected CIRs. For each event, we derive the average polytropic index, $\gamma_{av}$, of identified regions using the method outlined in Dayeh et al. (2022), with the steps summarized below and illustrated in Figure 2 using a selected event. Note that details of this method and its sensitivity have also been substantiated in previous studies (e.g., Kartalev et al., 2006; Livadiotis et al., 2019).

i. **Time interval selection (Figure 2a)**: We identify all sequences of 5-minute moving windows comprising five data points as was done in Dayeh et al. 2022. This condition minimizes mixed particle measurements across different streamlines.

ii. **Data filtration (Figure 2b)**: For each selected interval, we examine the stability of Bernoulli's integral (Livadiotis, 2016) by requiring the variance over the mean of the computed integral to be less than 10%. This condition further ensures that $\gamma$ is derived for plasma parcels likely along individual streamlines, where the polytropic relation is valid (Kartalev et al., 2006).

iii. **Density-Temperature fitting (Figure 2c)**: We perform a weighted linear fit to the plot of the natural logarithms of $T$ and $n$ (Equation 1; Figure 2b) to infer the value of $\gamma$, given by:

$$\ln T_p = (\gamma - 1) \cdot \ln n_p + constant$$

iv. **Determining γ (Figure 2d)**: Repeating step iii for all time windows in any region results in a set of polytropic indices, from which the mean $\gamma$ for each region of the CIR is then determined. The left panel in Figure 2d shows a 2D histogram of all $\gamma$ values corresponding to the uncompressed FSW region of Event 33 (1999-07-25 16:11 – 16:27 UT) as a function of solar wind speed, and the right plot shows the distribution of $\gamma$ within this region.

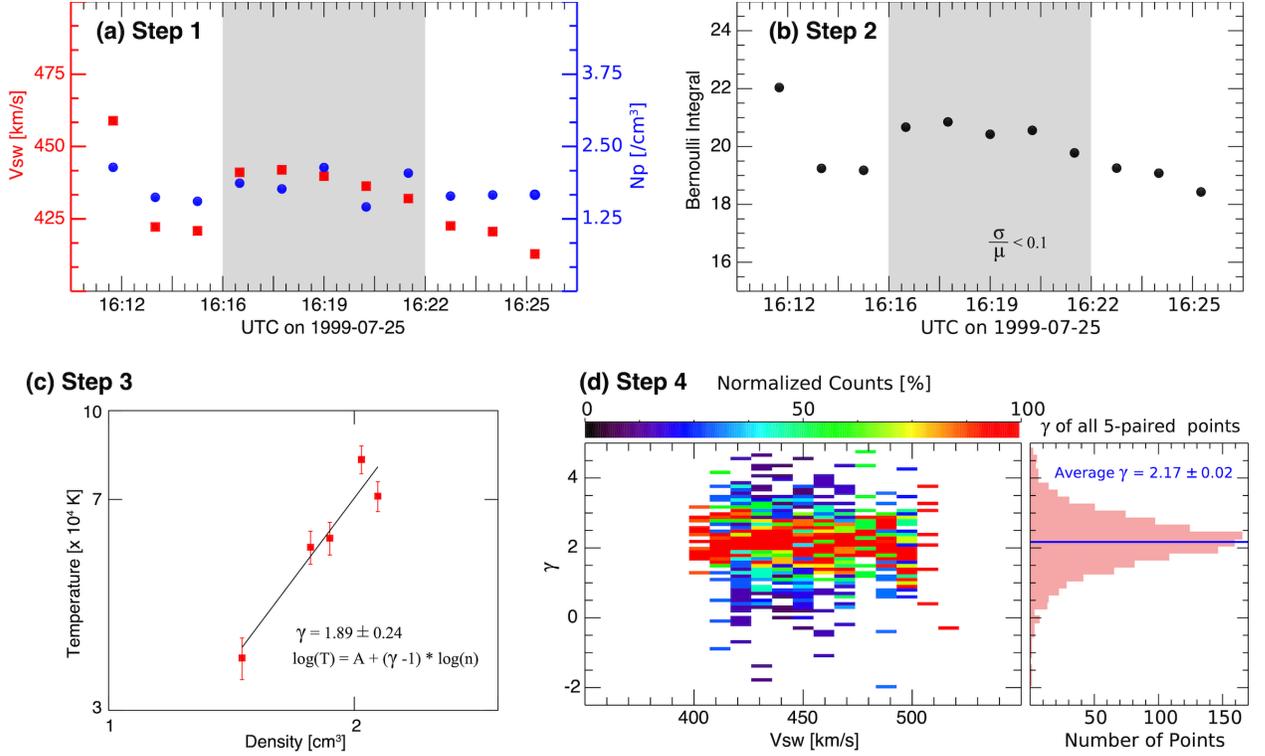

***Figure 2:*** *(a; Step 1) Solar wind speed (red) and proton density (blue) for an 11-minute period within the uncompressed FSW region of Event 33. (b; Step 2) Values of the Bernoulli integral for the corresponding data shown in (a). In both (a) and (b), the shaded interval illustrates a single 5-minute interval over which the stability condition for the Bernoulli integral must be met. (c; Step 3) Logarithmic plot of T versus n fit to a linear model (black line), from which γ is inferred. (d; Step 4; left) 2D histogram of the normalized number of the derived γ (count percentage) as a function of the solar wind speed. (d; Step 4; right) 1D histogram of the derived γ.*

**Calculating Average *γ* for Each Region**

To calculate the average *γ* for each region of a CIR, we first calculate *γ* as a function of solar wind speed by averaging all *γ* values within each solar wind speed bin (see 2D histogram in Figure 2). This is necessary to ensure that we describe a real thermodynamic process; as such, thermodynamics is independent of the comoving system and its kinetic characteristics (e.g., bulk speed). On the contrary, the presence of nonstationary effects, which disturb the polytropic processes and result to polytropic behavior having statistically significant dependence on bulk speed, would lead to unreliable results about the underlying polytropic process, the deduced polytropic index, and the thermodynamics, in general (Livadiotis & Desai 2016; Livadiotis 2018).

Next, we further constrain our event selection by requiring that all four characterized regions have derived indices that are statistically significant, and that the set of $\gamma$ values has no significant trend with speed (e.g., Livadiotis & Desai 2016; Dayeh et al. (2022).

This post selection criteria eliminated 31 additional events from our study, which resulted in a final list of 117 events. The average value of $\gamma$ for each region of each CIR event is then calculated as the weighted average normalized to the solar wind speed. This results in a single average $\gamma$ for each region of each CIR event included in this study.

**Results**

Figure 3a shows the histograms of the average polytropic index for all 117 events and within each region of the CIRs illustrated in Figure 1b. Figure 3a also displays the enriched histograms for the selected events, which are constructed to incorporate both the binned values and their associated uncertainties. Histograms of a data set are unreliable when the uncertainties are highly variable. To address this, illustrated enriched histograms are created. For each data point and its uncertainty, we generate a normally-distributed set of $10^3$ values, where the mean and standard deviation match the original value and its uncertainty. This method adjusts the distribution of the original values to better account for the uncertainties (for further details, see Livadiotis 2016, Dayeh & Livadiotis 2022).

Figure 3b shows the average value of $\gamma$ over all events as a function of region through the CIR, as determined from the enriched histograms. In the uncompressed SSW, the average value of $\gamma$ is near 5/3, which indicates that the plasma is nearly adiabatic. As the plasma becomes compressed, $\gamma_{avg}$ decreases which is indicative of energy transfer into the plasma via compressional heating; in particular, the polytropic index difference from its adiabatic value is proportional to the heating rate (Livadiotis 2018; Dayeh & Livadiotis 2022; Livadiotis & McComas 2023a; Katsavrias et al. 2024). In the compressed FSW, $\gamma_{avg}$ is back to the adiabatic value level, and lastly, in the uncompressed FSW, $\gamma_{avg}$ is well above the adiabatic value and above that of the compressed FSW. $\gamma$ values of the uncompressed FSW indicate heated plasma and is the most radiant among the four CIR regions studied here.

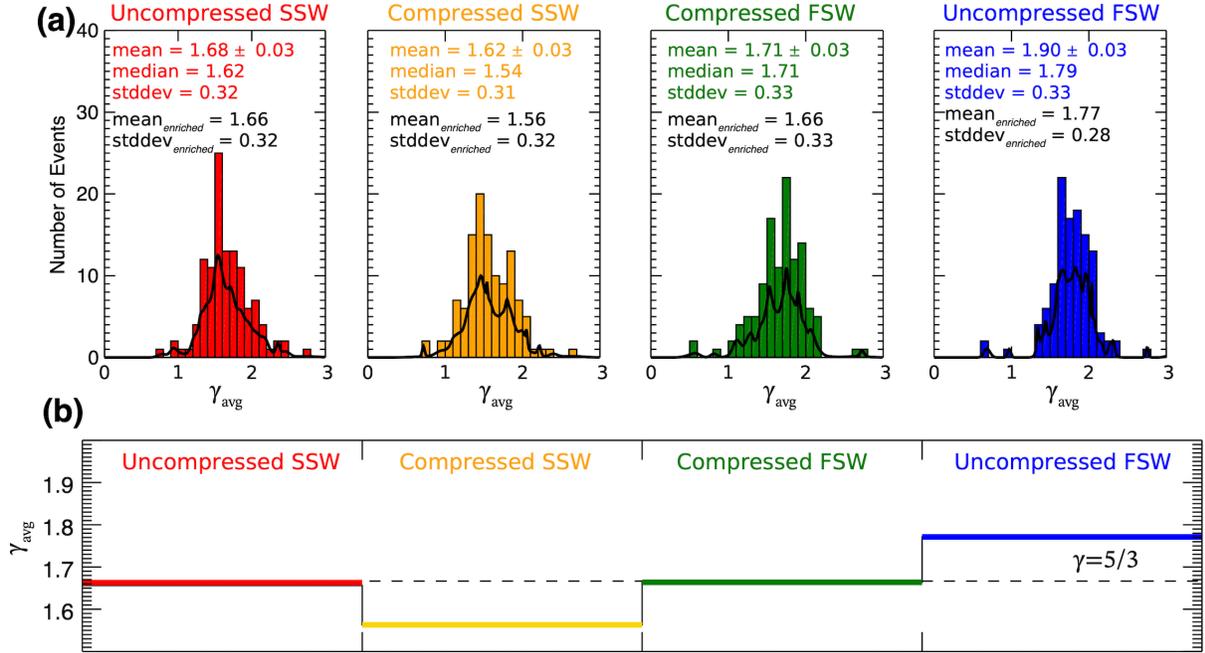

**Figure 3:** (a) Histograms of the average polytropic index for all events and each region of the CIR. Thick black lines indicate the enriched histograms used to calculate the mean value of $\gamma$. (b) Mean value of $\gamma_{avg}$ from the middle panels as a function of region of the CIR.

**Discussion and Conclusions**

To investigate the mechanism driving the polytropic index above the adiabatic value in the uncompressed FSW region, we analyzed Alfvénic wave activity within each CIR region.

We calculate the fluctuations in the magnetic field (B) and velocity (V) along different components of the GSE coordinate system. These fluctuations are determined by calculating the residual (deviations from the mean) by subtracting a running mean, computed using a 1-hour time window, from the respective data. Figure 4a shows the correlations between fluctuations in the solar wind velocity ($\delta V_i$) and magnetic field ($\delta B_i$) components in the X-direction of the GSE coordinate system within each region of Event 2. The correlation is observed to be strongest in the uncompressed FSW region, suggesting the presence of propagating Alfvén waves (e.g., Belcher and Davis, 1971). When this analysis was repeated for all events, it resulted in distributions of the correlation coefficients for each region. Figure 4b presents these results, summarizing the correlations across all events. As we transition from the uncompressed SSW to the uncompressed FSW, the data becomes highly correlated, which is a common trend for all 3 coordinate components (not shown) and observed across all events. The resulting distributions of the $X_{GSE}$ correlation coefficient across all events are then shown in Figure 4b.

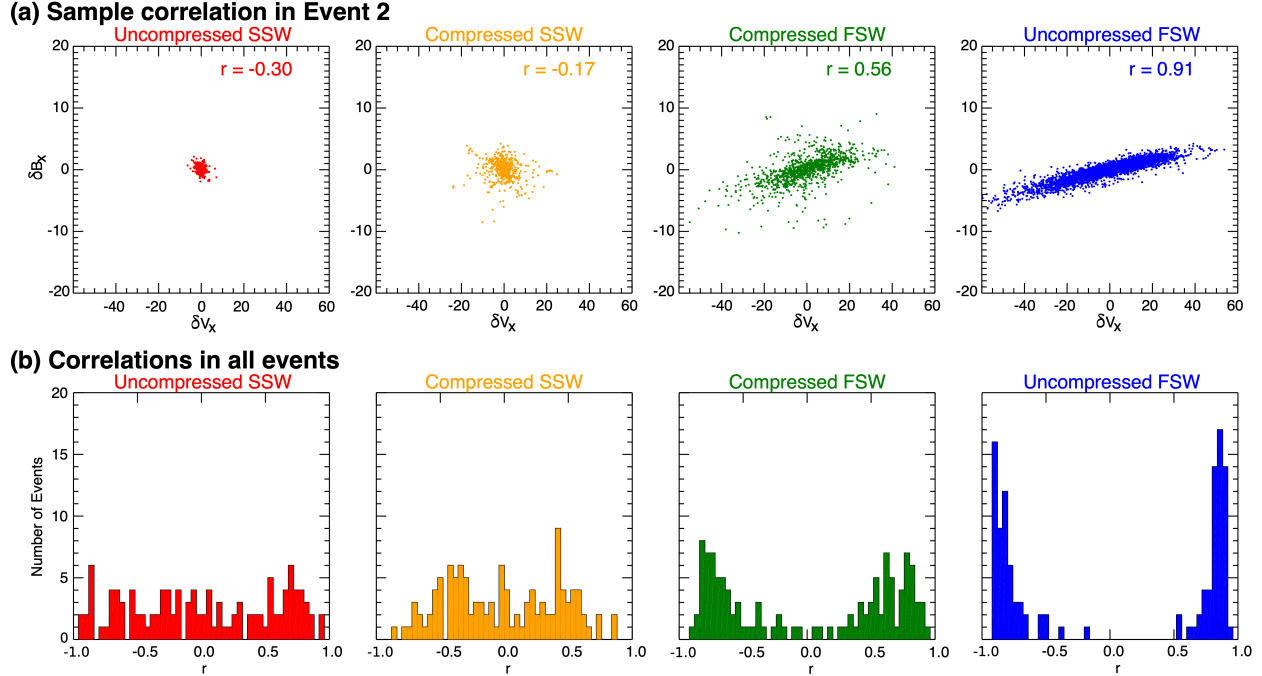

**Figure 4:** (a) Correlation of dB vs dV in the $X_{GSE}$ direction in Event 2 for the four CIR subregions. (b) Distributions of the correlation coefficients determined for all CIR events using analysis similar to that in panel (a).

An interesting trend prevails: The correlation distributions are observed to transition from nearly uniform in the uncompressed SSW to highly preferential towards +/- 1.0 in the uncompressed FSW. Note, that the uniform nature of the distribution in the SSW was validated by performing a sensitivity analysis of the duration of the upstream time window. Figure 4 shows results where the SSW region is 1 day in length, but the uniformity remained for 12 and 6 hour time windows. This high degree of correlation between the magnetic and velocity residual components in the uncompressed FSW is strongly indicative of Alfvénic wave activity (D'Amicis et al., 2022; Belcher, 1971). Thus, the increase in the value of the polytropic index appears to be driven by the increasing Alfvén wave activity in the uncompressed fast solar wind portion of the CIR. This component then radiates energy to the compressed FSW component.

The polytropic index characterizes compressions or expansions in the system and equivalently heat transfer processes. Although, dissipation of turbulence is not necessarily related to the polytropic process and is intrinsically different (Goldstein et al. 1995), Alfvénic activity associated in the FSW appears to be altering the polytropic index in the compressed slow

SW of CIR structures (Figure 4b). This observation provides evidence of slow SW heating by the fast SW in CIRs.

**Acknowledgements**

MAD acknowledges support from NASA LWS grants 80NSSC19K0079, 80NSSC21K1307, 80NSSC21K1316, and PSP GI award 80NSSC21K1769. MAD and GL acknowledge support from the IBEX mission under grants 80NSSC18K0237 and 80NSSC20K0719. RCA acknowledges support from NASA grants 80NSSC24K0908 and 80NSSC21K0733.